\documentclass[conference]{IEEEtran}
\usepackage{spconf,amsmath,graphicx}

\usepackage{enumitem}
\setlist{nosep, leftmargin=14pt}

\usepackage{mwe} 

\usepackage{cite}
\usepackage{amsmath,amssymb,amsfonts}
\usepackage{algorithmic}
\usepackage{graphicx}
\usepackage{textcomp}
\usepackage{graphicx}
\usepackage{amsmath}
\usepackage{hyperref}
\usepackage{graphicx}
\usepackage{subcaption}
\usepackage{float}
\usepackage{multirow} 
\usepackage{graphicx}
\usepackage{caption}
\usepackage{booktabs}

\title{PulmoFusion: Advancing Pulmonary Health with
Efficient Multi-Modal Fusion}
\name{Ahmed Sharshar$^*$, Yasser Attia$^{\dag}$, Mohammad Yaqub$^*$, Mohsen Guizani$^{\dag}$}
\address{$*$ Department of Computer Vision, MBZUAI, UAE\\
$\dag$ Department of Machine Learning, MBZUAI, UAE\\
\{ahmed.sharshar, yasser.attia, mohammad.yaqub, mohsen.guizani\}@mbzuai.ac.ae}
\begin{document}
%
\maketitle

\begin{abstract}
Traditional remote spirometry lacks the precision required for effective pulmonary monitoring. We present a novel, non-invasive approach using multimodal predictive models that integrate RGB or thermal video data with patient metadata. Our method leverages energy-efficient Spiking Neural Networks (SNNs) for the regression of Peak Expiratory Flow (PEF) and classification of Forced Expiratory Volume (FEV1) and Forced Vital Capacity (FVC), using lightweight CNNs to overcome SNN limitations in regression tasks. Multimodal data integration is improved with a Multi-Head Attention Layer, and we employ K-Fold validation and ensemble learning to boost robustness. Using thermal data, our SNN models achieve 92\% $\pm$ 2\% accuracy on a breathing-cycle basis and 99.5\% $\pm$ 0.5\% patient-wise. PEF regression models attain Relative RMSEs of 0.11 $\pm$ 0.05 (thermal) and 0.26 $\pm$ 0.07 (RGB), with an MAE of 4.52\% for FEV1/FVC predictions, establishing state-of-the-art performance.\footnote{Code and dataset can be found on \url{https://github.com/ahmed-sharshar/RespiroDynamics.git}}
\end{abstract}


\section{Introduction}

Asthma and Chronic Obstructive Pulmonary Disease (COPD) pose significant challenges to global health \cite{b1}, affecting approximately 339 million people and leading to over 3.91 million deaths annually, with COPD accounting for  6\% of global mortality \cite{b2}. ID-19 pandemic has highlighted the critical need for efficient and remote lung health assessment methods \cite{b5}. Traditional methods such as spirometry, which measures key parameters like Forced Vital Capacity (FVC), Forced Expiratory Volume in one second (FEV1), and Peak Expiratory Flow (PEF), are often limited by issues of cost, accessibility, and hygiene, particularly in low-resource environments \cite{b6}, prompting the exploration of non-invasive, continuous monitoring technologies through smartphones and wearable devices \cite{b60}. However, all the previous techniques lack generalization and fail to incorporate specific patient-related personal data. 


Recent innovations in this domain, such as thermal detection \cite{b17}, advanced PPG \cite{b18}, and optical imaging \cite{b25} have shown potential with reduced error rates, despite challenges like model overfitting and contrast issues \cite{b30}. Mobile applications like SpiroSmart and SpiroCall offer accessible lung function assessments but face robustness issues, with Mean Absolute Errors (MAEs) of 5.1\% and 7.2\%, respectively \cite{b33,b34}. Mobile thermal imaging respiratory oscillometry has shown notable accuracy in diagnosing pulmonary dysfunction, especially deep learning (DL) \cite{b38,b39}. UbiLung's passive monitoring approach has demonstrated efficacy in disease classification, achieving a 7.47\% MAE in FEV1/FVC ratio estimation \cite{b37}.  Although ideally, such solutions should run on a low resource setting, most DL-based solutions require high computational resources.

Spiking Neural Networks (SNNs) are bio-inspired, energy-efficient neural networks that process temporal data by mimicking efficiently the human brain. Their use extends beyond healthcare to fields like biometric security and brain-inspired computing \cite{b41}. SNNs' application in medical data analysis underscores their potential to revolutionize medical diagnostics in low resource settings\cite{b48}.

\par
This paper proposes an innovative approach to lung health assessment, leveraging either RGB or thermal video data with patient metadata, including height, age, athletic activity, and smoking status, to enhance model accuracy. Central to our approach is the innovative use of SNNs and the video-efficient model, highlighting their versatility and energy efficiency in healthcare. Furthermore, we conduct a comparative analysis of SNN- and CNN-based models, focusing on their accuracy and efficiency performance for classification and regression. Our contributions are:
\begin{itemize}
   \item We introduce PulmoFusion, an end-to-end lung health assessment model utilizing regression and classification. This approach incorporates data augmentation, multi-head attention, and ensemble learning.
   \item To the best of our knowledge, we are the first to introduce SNN to analyze thermal videos and efficiently integrate multi-modal thermal or RGB videos along with metadata for lung health assessment.
    \item We achieved state-of-the-art performance on FEV1/FVC metrics using our multimodal approach.
\end{itemize}

\begin{figure}[t]
\centering
\includegraphics[width=0.48\textwidth]{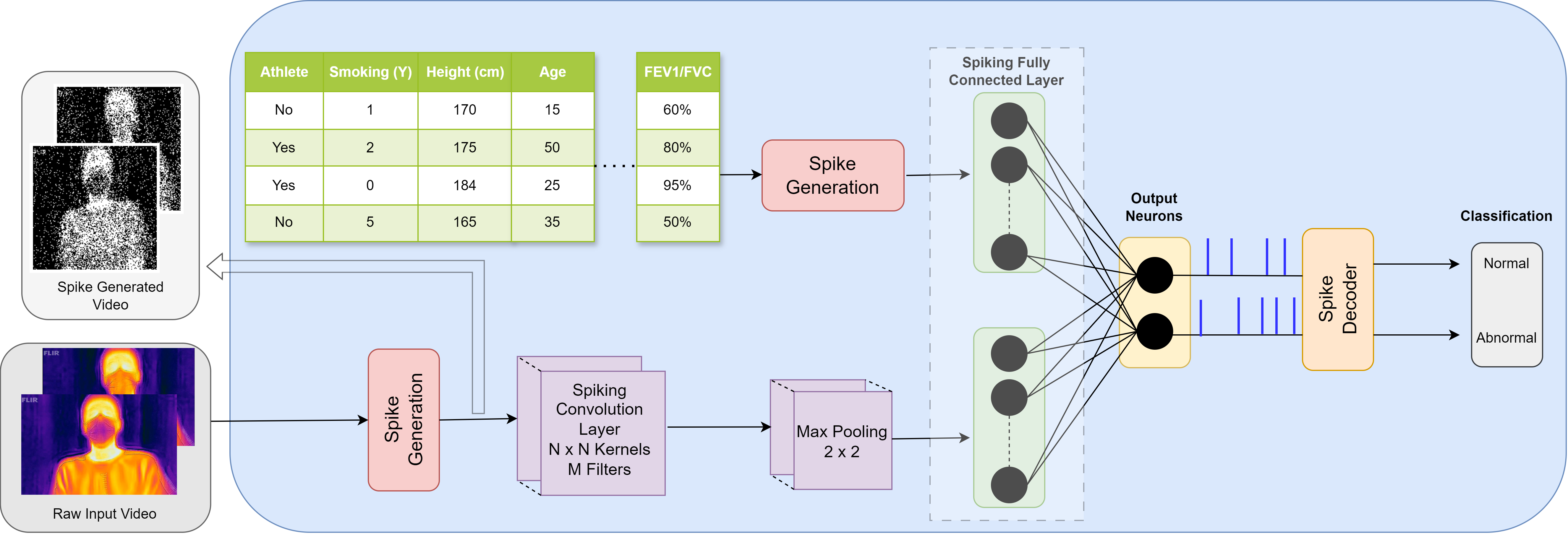}
\caption{PulmoFusion SNN Architecture: Processes thermal/RGB videos and metadata by encoding them into spikes. Video spikes feed into a Spiking CNN, undergo max pooling, and pass through a fully connected (FC) layer. Features from videos and metadata's FC layer are concatenated and forwarded to the classifier, which classifies if the video is normal if FEV1/FVC $\geq$ 70\% or abnormal if less.}
\label{fig: snn arch}
\end{figure}

\section{Methodology}
We propose classification and regression models to assess lung health accurately. In addressing the complexity problem, where the measurements of an entire breathing-blow cycle are crucial, our approach necessitates a shift from conventional image models to a video-based learning model. Therefore, X3D model is used as our vanilla (base) model.

\subsection{Spiking Neural Networks (SNNs) Models}

SNNs excel in energy efficiency and temporal precision by processing data with spike timing, closely modeling biological neurons. Central to our SNN model is the Leaky Integrate-and-Fire (LIF) neuron, which dynamically updates its membrane potential \(m^{(\ell)}(t)\) at any time \(t\) in layer \(\ell\) as:
\begin{equation}
    m^{(\ell)}(t) = \beta v^{(\ell)}(t - 1) + W^{(\ell)} s^{(\ell-1)}(t),
\end{equation}
where \(W^{(\ell)}\) are learnable weights, \(\beta\) is the decay rate, \(s^{(\ell-1)}(t)\) represents input spikes from the previous layer, and \(v^{(\ell)}(t)\) is the membrane potential at \(t\). If \(m^{(\ell)}(t)\) surpasses a threshold \(V_{\text{th}}^{(\ell)}\), a spike \(s^{(\ell)}(t)\) is emitted:
\begin{equation}
    s^{(\ell)}(t) = H(m^{(\ell)}(t) - V_{\text{th}}^{(\ell)}) =
    \begin{cases} 
        1, & \text{if } m^{(\ell)}(t) \geq V_{\text{th}}^{(\ell)}, \\
        0, & \text{otherwise}.
    \end{cases}
\end{equation}
After firing, the neuron’s membrane potential resets:
\begin{equation}
    v^{(\ell)}(t) = m^{(\ell)}(t) - s^{(\ell)}(t) V_{\text{th}}^{(\ell)}.
\end{equation}
To address challenges in gradient-based optimization due to non-differentiable spikes, we apply a surrogate gradient that approximates \(H'\) with a differentiable function, enhancing backpropagation efficiency. Despite recent advances, SNNs remain limited for regression tasks, especially in critical applications like healthcare. Thus, we focus on classification tasks where SNNs are more reliable.

Our PulmoFusion Convolutional SNN (CSNN) model processes video sequences to identify spatiotemporal patterns. Frames are normalized and rate-coded into binary spike trains \(u_{ij} \sim \text{Bernoulli}(1, p_{ij})\), where \(p_{ij}\) depends on pixel intensity. This rate-coded data flows through a 2D spiking convolutional layer, pooling, and a flat layer, with final classification based on spike frequency. In our multimodal approach, video data and metadata are converted into spike trains and fused through a fully connected layer, allowing comprehensive feature integration for enhanced analysis.

For SNN models, both thermal and RGB model architectures are used as in Figure \ref{fig: snn arch} with one multi-modal for thermal, while RGB needs ensemble models, consisting of four models with diverse convolutional filters and parameters. This ensemble methodology effectively balanced computational efficiency with predictive performance, significantly enhancing diagnostic accuracy. 

\begin{figure}[t]
\centering
\includegraphics[width=0.48\textwidth]{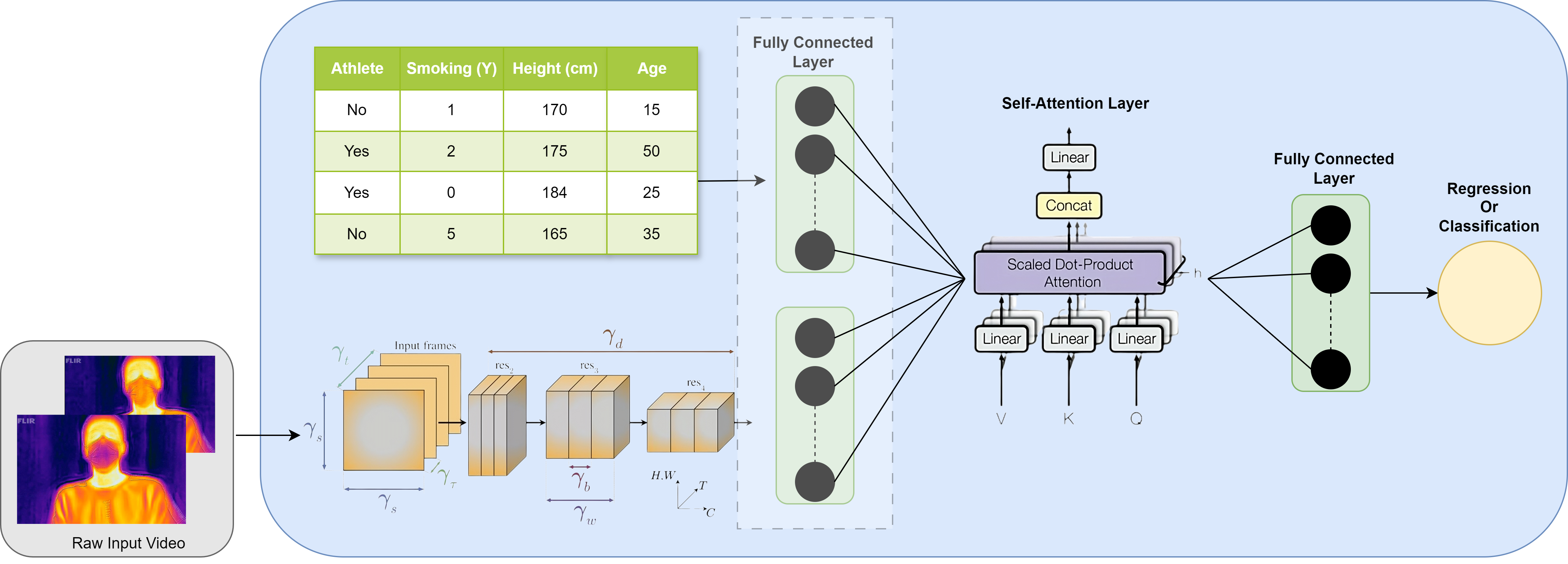}
\caption{PulmoFusion CNN Architecture: Processes thermal/RGB videos using X3D model. The model analyses the input videos as packets and output features. Features from the metadata are extracted using a fully FC layer. Then features from metadata and videos are concatenated using the attention layer and then the FC layer which is used as regression for PEF and FEV1/FVC values or classification.}
\label{fig: cnn arch}
\end{figure}

\begin{table*}[t!]
\caption{Results ($\mu$ $\pm$ $\sigma$) of the Single and Multi-Modal Analysis for Thermal and RGB Classification  Models.\label{classification}}
\centering
\small
\resizebox{0.70\textwidth}{!}{%
\begin{tabular}{llcccc}
\hline
\textbf{Model Type} & \textbf{Model Details} & \textbf{Recall $\uparrow$} & \textbf{Precision $\uparrow$} & \textbf{F1-Score $\uparrow$} & \textbf{Accuracy $\uparrow$} \\
\hline
\multicolumn{6}{c}{\textbf{Single Model}} \\
\hline
\multirow{4}{*}{SNN} & Thermal Model-Breathing Cycles & 84.15 \(\pm 1.5\%\) & 88.56 \(\pm 2.5\%\) & 86.03 \(\pm 3.0\%\) & 88.03 \(\pm 2.0\%\) \\
& \textbf{\textit{Thermal Model-Patient Wise}} & \textit{\textbf{94.12 \(\pm 0.5\%\)}} & \textit{\textbf{93.10 \(\pm 1.0\%\)}} & \textit{\textbf{93.91 \(\pm 0.6\%\)}} & \textit{\textbf{94.25 \(\pm 0.9\%\)}}\\
& RGB Model-Breathing Cycles & 75.26 \(\pm 2.5\%\) & 77.71 \(\pm 1.5\%\) & 76.47 \(\pm 3.0\%\) & 76.77 \(\pm 2.0\%\) \\
& RGB Model-Patient Wise & 92.77 \(\pm 1.0\%\) & 92.07 \(\pm 2.0\%\) & 92.92 \(\pm 1.0\%\) & 92.28 \(\pm 1.5\%\)\\
\hline
\multirow{4}{*}{CNN} & Thermal Model-Breathing Cycles & 87.13 \(\pm 2.0\%\)& 90.39 \(\pm 1.5\%\) & 89.51 \(\pm 2.5\%\) & 89.99 \(\pm 2.0\%\) \\
& \textbf{\textit{Thermal Model-Patient Wise}} & \textit{\textbf{94.25 \(\pm 0.5\%\)}} & \textit{\textbf{94.00 \(\pm 1.0\%\)}} & \textit{\textbf{94.12 \(\pm 0.8\%\)}} & \textit{\textbf{94.50 \(\pm 0.5\%\)}}\\
& RGB Model-Breathing Cycles & 78.85 \(\pm 1.5\%\)& 81.31 \(\pm 2.0\%\)& 79.04 \(\pm 2.0\%\)& 81.70 \(\pm 1.0\%\)\\
& RGB Model-Patient Wise & 94.05 \(\pm 2.0\%\) & 94.62 \(\pm 1.5\%\) & 94.25 \(\pm 1.0\%\) & 94.12 \(\pm 1.5\%\) \\
\hline
\multicolumn{6}{c}{\textbf{Multi-Modal}} \\
\hline
\multirow{4}{*}{SNN} & Thermal Model-Breathing Cycles & 89.04 \(\pm 2.0\%\) & 93.25 \(\pm 1.0\%\) & 91.61 \(\pm 3.0\%\) & 91.99 \(\pm 2.0\%\) \\
& \textbf{\textit{Thermal Model-Patient Wise}} & \textit{\textbf{99.00 \(\pm 0.4\%\)}} & \textit{\textbf{98.00 \(\pm 2.0\%\)}} & \textit{\textbf{98.80 \(\pm 0.5\%\)}} & \textit{\textbf{98.50 \(\pm 0.5\%\)}}\\
& RGB Model-Breathing Cycles & 79.95 \(\pm 3.0\%\) & 82.23 \(\pm 1.0\%\) & 81.07 \(\pm 2.0\%\) & 81.17 \(\pm 1.0\%\) \\
& RGB Model-Patient Wise & 98.50 \(\pm 1.5\%\) & 98.60 \(\pm 1.0\%\) & 98.44 \(\pm 1.3\%\) & 99.00 \(\pm 1.0\%\)\\
\hline
\multirow{4}{*}{CNN} & Thermal Model-Breathing Cycles & 91.87 \(\pm 2.0\%\)& 95.47 \(\pm 1.5\%\) & 94.37 \(\pm 2.5\%\) & 94.89 \(\pm 2.0\%\) \\
& \textbf{\textit{Thermal Model-Patient Wise}} & \textit{\textbf{99.50 \(\pm 0.5\%\)}} & \textit{\textbf{99.00 \(\pm 1.0\%\)}} & \textit{\textbf{99.20 \(\pm 0.8\%\)}} & \textit{\textbf{99.50 \(\pm 0.5\%\)}}\\
& RGB Model-Breathing Cycles & 82.90 \(\pm 2.0\%\)& 85.59 \(\pm 1.5\%\)& 83.10 \(\pm 3.0\%\)& 86.00 \(\pm 1.5\%\)\\
& RGB Model-Patient Wise & 99.00 \(\pm 1.0\%\) & 99.50 \(\pm 0.5\%\) & 99.00 \(\pm 1.0\%\) & 99.30 \(\pm 0.2\%\) \\
\hline
\end{tabular}}
\end{table*}

\subsection{CNN Model- Unimodel and Multimodel}
We investigate the X3D model small \cite{b52} for CNN regression and classification due to its ability to extend 2D images to handle spatial, temporal, width, and depth expansions. The model's expansion strategy includes adjustments in temporal duration ($\gamma : t$), frame rate ($\gamma : \tau$), spatial resolution ($\gamma : s$), network width ($\gamma : w$), bottleneck width ($\gamma : b$), and depth ($\gamma : d$), enabling computational effective 3D video analysis.

We developed models for distinct data types: thermal and RGB videos, finding solid correlations between participant metadata—such as smoking habits, age, height, and athletic status—and pulmonary health. Building on these insights, we designed a multimodal model that integrates video data with metadata, enhancing predictive accuracy.

To optimize data fusion, we tested two approaches: a basic Dense Layer for straightforward integration and a Multi-Head Attention Layer for more nuanced merging of CNN and metadata features. This attention mechanism allows the model to focus on critical features and deeper correlations.

Our model, illustrated in Figure \ref{fig: cnn arch}, uses an X3D CNN backbone to extract spatial-temporal features from videos. In contrast, metadata features are processed through a fully connected (FC) layer to prevent overfitting. The Multi-Head Attention Layer fuses these features, fed into a final FC layer for either regression or classification tasks.

Furthermore, we employ ensemble learning techniques, utilizing multiple models to improve learning. These techniques offer several advantages, such as increased robustness against overfitting, adept handling of non-linear relationships, and notably enhanced prediction accuracy.

\section{Dataset}

To overcome the lack of availability of multi-modalities datasets for lung health assessment, we collected a novel dataset assembled using 60 volunteers \cite{res}. These participants provided a wide range of data in terms of age (15-75 years), weight (51.2-102.7 kg), and height (154-189 cm), contributing to the dataset's diversity. They underwent two sessions: a resting state and a post-exercise state, each for 1 minute, generating 120 data points per type. The dataset contains RGB and thermal videos, heart rate, smartwatch electrocardiogram (ECG), blood pressure, and Peak Flow \& Asthma Meter, which we used as our ground truth values. The dataset also includes detailed metadata with personal and health-related information such as age, height, smoking duration, athlete status, seasonal cough indicator, lung past problems, lung genetic problems, and inbody data. 

To enhance lung health analysis, estimated values for FVC and FEV1 were included and calculated using NHANES III study reference equations\cite{b55}.  These values are used to determine if the person was healthy or not. If the FEV1/FVC percentage for the measured (using flow meter during data collection) over the predicted one (from the equations) \>70\% it means the subject is normal; otherwise, it implies there is an abnormality. This is a standard measurement by the American Lung Association \cite{b57}. This standard processing and value is used clinically to determine the patient's health status. PEF values were derived from the Nunn and Gregg equation\cite{b56}, which are compared with the measured ones to detect how healthy the participant is. The dataset was collected under the supervision of expert doctors. 

The experimental protocol ensured data integrity and consistency through a two-phase collection process, with steps for vital signs measurement, smartwatch  ECG recording, respiratory flow assessment, and simultaneous thermal and RGB video recording. Video synchronization was achieved using a timestamp camera application. The post-processing stage involved meticulous manual inspection of videos, facial isolation using OpenCV for RGB and thermal recordings. The final dataset contained 2,424 segmented videos, where each segment is just one full breathing cycle.

\begin{table*}[t!]
\caption{Results ($\mu$ $\pm$ $\sigma$) for the Thermal and RGB Analysis of the PEF and FEV1/FVC Regression Models. \label{regression}}
\centering
\small
\resizebox{0.70\textwidth}{!}{%
\begin{tabular}{llccc}
\hline
\textbf{Model Type} & \textbf{Model Details} & \textbf{Relative RMSE $\downarrow$} & \textbf{Relative MAE $\downarrow$} & \textbf{Pearson Correlation $\uparrow$} \\
\hline
\multicolumn{5}{c}{\textbf{PEF Regression}} \\
\hline
\multirow{5}{*}{Thermal Model} & Single Model (Baseline) & 0.30 $\pm$ 0.08 & 0.26 $\pm$ 0.07 & 0.72 $\pm$ 0.08 \\
& Augmentations & 0.22 $\pm$ 0.06 & 0.19 $\pm$ 0.06 & 0.80 $\pm$ 0.07 \\
& Multi-Modal & 0.15 $\pm$ 0.05 & 0.12 $\pm$ 0.08 & 0.89 $\pm$ 0.06 \\
& \textbf{Multi-Head Attention} & \textbf{0.13 $\pm$ 0.01} & \textbf{0.10 $\pm$ 0.03} & \textbf{0.90 $\pm$ 0.02} \\
& \textbf{\textit{Ensemble Learning}} & \textbf{\textit{0.11 $\pm$ 0.05}} & \textbf{\textit{0.09 $\pm$ 0.06}} & \textbf{\textit{0.93 $\pm$ 0.04}} \\
\hline
\multirow{5}{*}{RGB Model} & Single Model (Baseline) & 0.43 $\pm$ 0.05 & 0.40 $\pm$ 0.08 & 0.60 $\pm$ 0.06 \\
& Augmentations & 0.40 $\pm$ 0.07 & 0.30 $\pm$ 0.07 & 0.63 $\pm$ 0.09 \\
& Multi-Modal & 0.32 $\pm$ 0.07 & 0.25 $\pm$ 0.03 & 0.75 $\pm$ 0.07 \\
& \textbf{Multi-Head Attention} & \textbf{0.28 $\pm$ 0.08} & \textbf{0.23 $\pm$ 0.05} & \textbf{0.77 $\pm$ 0.05} \\
& \textbf{\textit{Ensemble Learning}} & \textbf{\textit{0.26 $\pm$ 0.07}} & \textbf{\textit{0.21 $\pm$ 0.04}} & \textbf{\textit{0.79 $\pm$ 0.04}} \\
\hline
\multicolumn{5}{c}{\textbf{FEV1/FVC Regression}} \\
\hline
\multirow{5}{*}{Thermal Model} & Single Model (Baseline) & 0.19 $\pm$ 0.06 & 0.16 $\pm$ 0.08 & 0.85 $\pm$ 0.09 \\
& Augmentations & 0.17 $\pm$ 0.04 & 0.14 $\pm$ 0.05 & 0.88 $\pm$ 0.05 \\
& Multi-Modal & 0.12 $\pm$ 0.04 & 0.10 $\pm$ 0.06 & 0.92 $\pm$ 0.02 \\
& \textbf{Multi-Head Attention} & \textbf{0.09 $\pm$ 0.03} & \textbf{0.08 $\pm$ 0.02} & \textbf{0.94 $\pm$ 0.02} \\
& \textbf{\textit{Ensemble Learning}} & \textbf{\textit{0.06 $\pm$ 0.03}} & \textbf{\textit{0.05 $\pm$ 0.05}} & \textbf{\textit{0.96 $\pm$ 0.01}} \\
\hline
\multirow{5}{*}{RGB Model} & Single Model (Baseline) & 0.33 $\pm$ 0.07 & 0.30 $\pm$ 0.05 & 0.71 $\pm$ 0.07 \\
& Augmentations & 0.30 $\pm$ 0.06 & 0.27 $\pm$ 0.04 & 0.74 $\pm$ 0.06 \\
& Multi-Modal & 0.24 $\pm$ 0.05 & 0.22 $\pm$ 0.04 & 0.80 $\pm$ 0.04 \\
& \textbf{Multi-Head Attention} & \textbf{0.18 $\pm$ 0.04} & \textbf{0.16 $\pm$ 0.03} & \textbf{0.82 $\pm$ 0.03} \\
& \textbf{\textit{Ensemble Learning}} & \textbf{\textit{0.12 $\pm$ 0.03}} & \textbf{\textit{0.1 $\pm$ 0.04}} & \textbf{\textit{0.85 $\pm$ 0.05}} \\
\hline
\end{tabular}%
}
\end{table*}
\section{Experimental Setup}

Our experimental methodology was designed with two primary goals: regression, to estimate PEF and evaluate the FEV1/FVC ratio, and classification, to detect abnormalities using the FEV1/FVC ratio with a delineation threshold of 70\% for pulmonary dysfunction (38 subjects healthy, 22 not). We enhanced dataset generalization by allocating 80\% for training and 20\% for testing based on individual subjects. As the data size is not large, we utilized a pre-trined X3D model to finetune along with 5-fold cross-validation to ensure distinct subject sets across training and testing phases and spatial and temporal data augmentation. Video clips were standardized to 30 frames and resized to \(224 \times 224\) pixels.

The dataset included segmented video clips, each representing a unique respiratory cycle. Despite consistent metadata, we observed variability in respiratory performance. We implemented a post-processing technique to improve model accuracy by averaging respiratory metrics by participant to mitigate natural variability for the regression model. A majority voting scheme was implemented in the classification models, combining evaluations from sub-videos to determine a comprehensive diagnosis for each subject.

For SNN, models were optimized using ADAM optimizer and MSE count loss while CNN used ADAM and MSE loss for regression and categorical cross entropy for classification. The detailed hyperparameters and configurations for our SNN and CNN models will be available in the repository upon acceptance. Computational requirements for training varied between models: the CNN model demanded a single Quadro RTX 6000 GPU with 24 GB RAM, while the SNN model was more resource-efficient, requiring only 4 GB.

\section{Results \& Discussion }

Table \ref{classification} presents a detailed comparison between thermal and RGB classification models, highlighting the thermal models' superior accuracy per breathing cycle. However, both models enhanced across all metrics upon patient-wise aggregation, illustrating the significant role of data aggregation. While multi-modal CNNs outperform SNNs in initial performance, this gap diminishes considerably on a patient-wise level, showcasing comparable effectiveness. SNNs distinguish themselves with faster inference, processing patient data in just 0.2 seconds versus CNN's 1.3 seconds, underscoring SNNs' advantage for rapid processing.

The RGB model's accuracy for abnormal and normal conditions was 80.41\% and 83.23\% breathing-cycle-wise, respectively, which surged to 100\% patient-wise. The Thermal model showed a similar trend, starting from 90.72\% (abnormal) and 94.75\% (normal) to reaching 100\% accuracy for patient-wise. Notably, the thermal data outperformed RGB due to its ability to capture changes in mask heat distribution, reflecting variations in exhaled air volume.

Table \ref{regression} delineates the performance metrics of our regression models utilizing thermal and RGB imaging data, including Relative RMSE, Relative MAE, and Pearson Correlation coefficients. Our initial single-model efforts, leveraging video data, established a baseline with a Relative RMSE of 0.30 and a Pearson Correlation of 0.72. Despite demonstrating reasonable correlation, these results underscored the necessity for more accurately capturing video data.

To advance our model's performance, we employed techniques such as data augmentation to diminish error rates by diversifying the dataset and enhancing its generalization ability. Adopting a multi-modal incorporating metadata significantly enhanced the model's input. By implementing multi-head attention mechanisms, we enabled the model to concentrate precisely on relevant data features, thereby improving its ability to recognize complex patterns. 

We achieved the best results by adding ensemble learning to the multi-modal with Multi-Head Attention multi, achieving notable results with a Relative RMSE of $0.13$. We experimented with the best model for FEV1/FVC prediction and achieved MAE of 4.52\%, representing state-of-the-art results. Our model's robustness ensures superior generalization capacity, making it highly applicable.

\section{Conclusion}
We proposed a multi-modal, non-invasive pulmonary health evaluation approach that integrates RGB or thermal videos with patient metadata to assess lung function. By combining the complementary strengths of SNNs and CNNs within a unified framework, our methodology achieved notable advancements in accuracy, efficiency, and energy consumption. The results highlight the potential of thermal imaging, offering more precise insights into respiratory patterns. However, the reliance on high-quality, manually segmented datasets remains a critical bottleneck, potentially limiting scalability and real-world applicability. While our small participant pool and the scarcity of SNN regression models restrict generalizability, these limitations underscore the need for automated data preprocessing techniques, larger datasets, and further exploration of SNNs in regression tasks. Future work addressing these issues can help unlock the broader potential of this approach, making it more practical for widespread adoption.

\section{COMPLIANCE WITH ETHICAL STANDARDS}

This work involved human subjects or animals in its research. This study was performed in line with the principles of the Declaration of Helsinki. Approval of all ethical and experimental procedures and protocols was
granted by a written consent form signed by each participant for data usage for scientific research purposes only.

\section{ACKNOWLEDGMENT}
No funding was received for conducting this study. The authors have no relevant financial or non-financial interests to disclose.

\bibliographystyle{IEEEbib}
\bibliography{strings,refs}

\end{document}